\documentclass[aps,superscriptaddress,amsmath,amssymb,floatfix,twocolumn,showpacs,amsfonts,longbibliography]{revtex4-1}
\usepackage{times}
\usepackage[varg]{txfonts}
\usepackage{textcomp}
\usepackage{graphicx}
\usepackage{subfigure}
\usepackage{tabu}
\usepackage{color}
\usepackage[colorlinks=true,citecolor=blue,urlcolor=blue,linkcolor=blue,hyperindex]{hyperref}
\usepackage{braket}
\usepackage{overpic}
\usepackage{amssymb}

\allowdisplaybreaks

\begin{document}


\title{Indirect K-edge bimagnon resonant inelastic X-ray scattering spectrum of $\alpha$-FeTe}
\author{Zengye Huang}
\affiliation{State Key Laboratory of Optoelectronic Materials and Technologies, School of Physics, Sun Yat-Sen University, Guangzhou 510275, China}

\author{Sean Mongan}
\affiliation{Department of Chemistry and Physics, Augusta University, 1120 15$^{th}$ Street, Augusta, Georgia 30912, USA}

\author{Trinanjan Datta}
\email[Corresponding author:]{tdatta@augusta.edu}
\affiliation{Department of Chemistry and Physics, Augusta University, 1120 15$^{th}$ Street, Augusta, Georgia 30912, USA}
\affiliation{State Key Laboratory of Optoelectronic Materials and Technologies, School of Physics, Sun Yat-Sen University, Guangzhou 510275, China}

\author{Dao-Xin Yao}
\email[Corresponding author:]{yaodaox@mail.sysu.edu.cn}
\affiliation{State Key Laboratory of Optoelectronic Materials and Technologies, School of Physics, Sun Yat-Sen University, Guangzhou 510275, China}

\date{\today}

\begin{abstract}
We calculate the K-edge {\it indirect} bimagnon resonant inelastic x-ray scattering (RIXS) intensity spectra of the bicollinear antiferromagnetic order known to occur in the $\alpha$-FeTe chalcogenide system. Utilizing linear spin wave theory for this large-S spin system we find that the bimagnon spectrum contains four scattering channels (two intraband and two interband). We find from our calculations that for suitable energy-momentum combination the RIXS spectra can exhibit a one-, two- or three- peak structure. The number of peaks provides a clue on the various bimagnon excitation processes that can be supported both in and within the acoustic and optical magnon branches of the bicollinear antiferromagnet. Unlike the RIXS response of the antiferromagnetic or the collinear antiferromagnetic spin ordering, the RIXS intensity spectrum of the bicollinear antiferromagnet does not vanish at the magnetic ordering wave vector $(\pi/2,-\pi/2)$. It is also sensitive to next-next nearest neighbor and biquadratic coupling interactions. Our predicted RIXS spectrum can be utilized to understand the role of multi-channel bimagnon spin excitations present in the $\alpha$-FeTe chalcogenide.
\begin{description}
\item[PACS number(s)] 78.70.Ck, 75.25.-j, 75.10.Jm
\end{description}
\end{abstract}
%
%
%
%
%
\maketitle
\section{Introduction}
Over the past decade there has been a flurry of intense research activities to understand the underlying physical properties of iron-based superconductors\cite{RevModPhys.87.855}. These superconductors come with four flavors of crystal structure that can be generally be classified as belonging to the 1111 type RFeAsO (R represents rare earth elements), the 122 type Ba(Ca)Fe$_{2}$As$_{2}$, the 111 type LiFeAs, and the 11 type $\alpha$-FeTe~\cite{PhysRevLett.102.177003}. In contrast to the cuprates the magnetic ground state of the pnictides or chalcogenides can order in a wide variety of arrangements. For example, the LaFeAsO system and the  BaFe$_2$As$_2$~\cite{nature453.899,PhysRevLett.101.057003,PhysRevB.78.033111} compound display a collinear antiferromagnetic (CAF) phase, and the $\sqrt{5}\times\sqrt{5}$ block antiferromagnetic order (BAF)in K$_{0.8}$Fe$_{1.6}$Se$_{2}$   ~\cite{ChinPhysLett.28.086104}, and the semiconducting rhombus $\sqrt{5}\times\sqrt{5}$ iron vacancy ordered $245$ type in K$_{2}$Fe$_{4}$Se$_{5}$ ~\cite{PhysRevLett.109.267003} , and the bicollinear antiferromagnetic (BCAF) state in the ferrochalcogenide $\alpha$-FeTe systems~\cite{PhysRevLett.102.177003} .

Irrespective of the underlying pairing mechanism, arising either from a local moment or an itinerant electron based model, magnetism is known to play a crucial role in the physical description of iron-based superconductors ~\cite{srep00381,RevModPhys.84.1383}. The similarities between the nonmagnetic electronic band structure of of $\alpha$-FeTe, LaFeAsO, and BaFe$_{2}$As$_{2}$ hinted to the possibility that the magnetic ground state of all these materials would be of the CAF type. However, density functional theory calculations by Ma et al revealed a surprise! The system actually has BCAF ordering as its magnetic ground state~\cite{PhysRevLett.102.177003}.

Spin wave excitations in a magnetic system are typically measured with neutron scattering. In the iron pnictides spin wave dispersion measured by neutron scattering become diffusive with increasing energy, but, is still well defined up to 214 meV~\cite{NaturePhysics5.555,PhysRevB.84.054544}. The situation in the iron chalcogenide system $\alpha$-FeTe is rather different. The spectra becomes diffuse for energies above 85 meV without any well-defined excitations\cite{PhysRevLett.106.057004}. Experimentally it is difficult to detect the higher energy excitations for $\alpha$-FeTe. Thus, there is a possibility that the excitations in this range maybe composed of a mixture of single and multi-magnon, such as a bimagnon, excitation states .

In order to understand the contribution of magnetism to superconductivity we may need to probe the details of the high energy excitations~\cite{srep00381}. The presence of multi-magnon excitations could be potentially be probed via resonant inelastic X-ray scattering (RIXS) spectroscopy. Current experimental ~\cite{PhysRevLett.100.097001,PhysRevLett.102.167401,PhysRevLett.103.047401,PhysRevLett.104.077002,PhysRevLett.105.157006,PhysRevLett.107.107402,
NatPhys.7.725,PhysRevLett.108.177003,PhysRevB.85.214527,NatMater.11.850,NatMater.12.1019,PhysRevLett.103.107205,Nature.485.82,PhysRevLett.110.265502,PhysRevLett.110.087403,PhysRevB.95.235114}, theoretical~\cite{EuroPhysLett.80.47003,PhysRevLett.101.106406, PhysRevB.75.214414,PhysRevB.77.134428,NJP.11.113038,PhysRevB.89.165103,PhysRevB.92.035109,PhysRevLett.109.117401,PhysRevLett.110.117005,PhysRevLett.106.157205,PhysRevB.86.125103}, and computational studies ~\cite{PhysRevB.82.064513,ncomms2428,NJP.11.113038,PhysRevLett.106.157205,PhysRevB.83.245133,PhysRevLett.110.265502} on RIXS suggest the possibility to explore quantum magnets and correlated systems from various perspectives.

First-principles electronic structure calculations on LaFeAsO, confirmed by inelastic neutron scattering experiments, predicted a CAF phase and a semi-metallic state~\cite{PhysRevB.78.033111,nature453.899}. Existing studies suggest long range antiferromagnetic (AF) order due to the presence of superexchange interactions~\cite{PhysRevB.78.224517}. It has also been proposed that nesting between electron and hole Fermi surfaces can led to AF ordering~\cite{EuroPhysLett.83.27006,PhysRevLett.101.057003}. But, for the $\alpha$-FeTe system with diagonal AF order the Fermi surface nesting mechanism does not hold~\cite{PhysRevLett.102.177003}. Furthermore, density functional calculations on $\alpha$-FeTe predict magnetic moments in the 2.2$\mu_{B}$ - 2.6$\mu_{B}$ range. Thus, the BCAF ground state of $\alpha$-FeTe can be adequately captured using the $J_1-J_2-J_3-K$ Heisenberg model~\cite{PhysRevLett.102.177003,PhysRevB.85.144403}.

Resonant inelastic X-ray scattering spectroscopy offers an alternate way to measure high energy spin excitations in cuprates and pnictides~\cite{NaturePhysics7.725,ncomms2428,PhysRevB.84.020511}. While RIXS at the Fe L$_3$-edge can be used to measure single magnon excitations, as in neutron scattering, at the K-edge RIXS can probe bimagnon (multi-magnon) excitations. Recent comprehensive theoretical investigations on the K-edge bimagnon RIXS in the AF and the CAF phases have shown that the spectrum vanishes at the magnetic ordering vector for the AF and the CAF phases~\cite{PhysRevB.75.214414,PhysRevB.77.134428,PhysRevB.89.165103}. The RIXS spectra show a split-peak structure as spatial anisotropy or frustrating further neighbor interactions are tuned. The bimagnon excitations involve a local modification of the superexchange interaction mediated via the core hole, thus leading to the RIXS spectra expressed as a momentum-dependent four-spin correlation function~\cite{EuroPhysLett.80.47003,PhysRevB.77.134428, EuroPhysLett.73.121}.

In this article we investigate the {\it indirect} K-edge bimagnon RIXS spectrum of the BCAF ordered phase. Our goal is to develop an understanding of the RIXS spectrum features of this multi-band system with further neighbor and biquadratic coupling interactions. Theoretical analysis of a multi-channel RIXS spectrum at the K-edge in a chalcogenide is missing. In this regard the $\alpha$-FeTe system provides a canonical example of a multi-band system that can host multi-magnon (bimagnon) excitations scattering via multiple RIXS channels, at the linear spin wave theory level (LSWT). The above reasoning provides the primary reason behind pursuing the calculation, in addition to offering clues on potential high energy excitations hosted in chalcogenides.

We find that the RIXS spectra of the BCAF phase contains four channels of scattering, two of which are classified as intraband and the other two as interband. The two magnon branches create acoustic and optical magnons that can scatter between themselves to produce the multi-channel RIXS spectra. In contrast to the AF and CAF phase K-edge RIXS response which vanishes at the ordering wave vector, in the BCAF phase the intensity is non-zero~\cite{PhysRevB.75.214414,PhysRevB.89.165103} . The RIXS spectrum develops several peaks originating in the intra- and inter- band scattering channels. For certain momentum choices along the magnetic Brillouin zone path the intra- or inter- channels may either contribute or be suppressed. This leads to the formation of a single [${\bf q}=(\pi,-\pi)$], a double [${\bf q}=(\pi,0)$], or a triple peak [${\bf q}=(\pi/2,-\pi/2)$] structure. Thus, our spin wave theory calculation captures several interesting and significant features of the RIXS response of the BCAF phase which in turn can provide more details on high energy multi-magnon excitations. We also find that the RIXS spectrum is affected by next-next nearest neighbor interactions and biquadratic couplings. In both, cases the sensitivity of the magnon bands to the parameters of the $\alpha$-FeTe model manifests itself in the RIXS spectrum. Finally, we note that  $\alpha$-FeTe is a large-S spin system. Thus, within spin-wave theory where quantum fluctuations are scaled by 1/S, we do not expect our linear spin wave theory results to be modified much. It also justifies neglecting the higher spin-spin interactions in the bimagnon channels. \cite{PhysRevB.79.054503,PhysRevB.89.165103}.

\begin{figure}[t]
\centering
{\subfigure[]{
\includegraphics[scale=0.5]{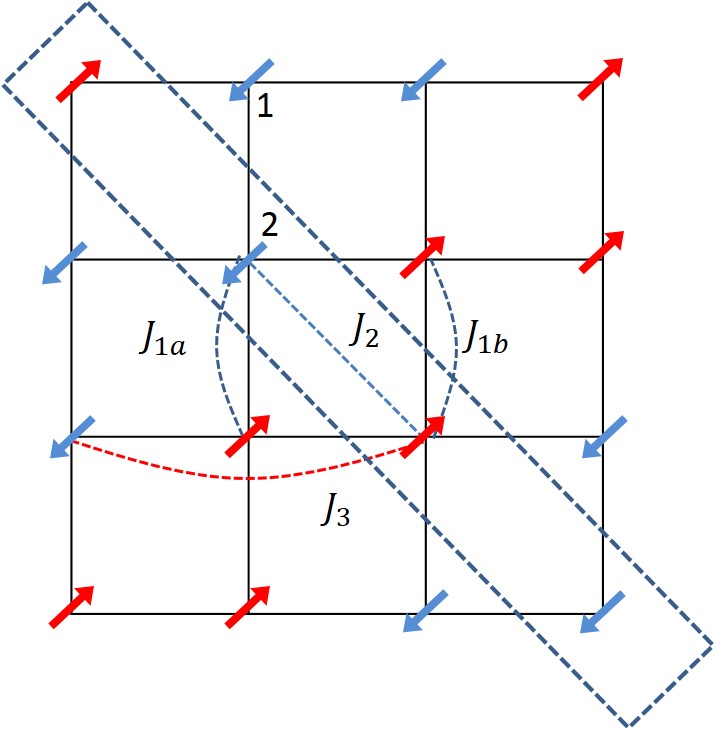}}}
{\subfigure[  ]{
\includegraphics[scale=0.5]{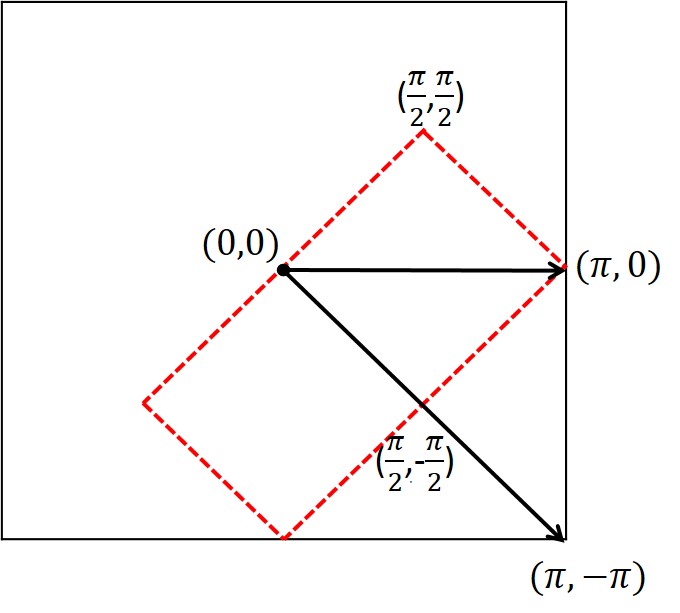}}}
\caption{Magnetic interactions in real space and magnetic Brillouin zone. (a) Schematic representation of the bicollinear antiferromagnetic (BCAF) model. $J_{1a}$ and $J_{1b}$ represent the coupling of two nearest-neighbor parallel and antiparallel spins, respectively. $J_2$ and $J_3$ are isotropic  $2^{nd}$ and $3^{rd}$ nearest-neighbor couplings, respectively. (b) Magnetic Brillouin zone of the BCAF model shown by the red dashed line. }
\label{fig:1}
\end{figure}

\section{Model}
\subsection{Heisenberg Model of Bicollinear Antiferromagnetic Phase}
Experimental investigation and first-principles calculation on the $\alpha$-FeTe system suggest that it is a semimetal with a strong magnetic moment of approximately 2.0 $\mu_{B}$  -2.5 $\mu_{B}$ around the Fe$^{2+}$ ion \cite{PhysRevLett.102.177003,PhysRevB.79.054503}. The presence of the large local moment allows one to describe the possible magnetic phases within a frustrated Heisenberg model, ignoring any weak itinerant effects that may be present.A unified minimum effective model that can capture the physics of the iron-based superconductors, including the BCAF ground state of the $\alpha$-FeTe system is given by
\begin{eqnarray}
\label{eq:ham}
 \mathcal{H}=\sum_{ij}[J_{ij}{\bf S}_i\cdot {\bf S}_j+K_{ij}({\bf S}_i\cdot {\bf S}_j)^{2}] ,\label{eq1}
\end{eqnarray}
where the nearest-neighbor (NN), the next nearest-neighbor (NNN), and the next-next nearest-neighbor (NNNN) magnetic exchange interactions are denoted by the symbols $J_1$,$J_2$, and  $J_3$. Neutron scattering studies on the chalcogenide system shows the existence of highly anisotropic NN exchange couplings. To model this behavior a non-Heisenberg biquadratic couplings, $K_{ij}=K$, is introduced if and only if $i,j$ are two nearest sites. As shown in figure~\ref{fig:1}(a) the BCAF model can be simplified to an effective $J_1-J_2-J_3-K$ model. The NN exchange couplings redefined depending on the alignment of the two spins. If the spins are antiparallel then $J_1a=J_1+2KS^2$. If they are parallel then $J_1b=J_1-2KS^2$. In equation (\ref{eq:ham}) the $J_1$ coupling is ferromagnetic (FM). But, $J_2$ and $J_3$ are antiferromagnetic \cite{PhysRevLett.102.177003, EuroPhysLett.86.67005, PhysRevB.85.144403}. The general model Hamiltonian above can support a host of different magnetic phases including AF, CAF, BCAF, and BAF phases. The BCAF phase is stable when $2J_3 > J_2$ and $2J_2 > |J_1|$~\cite{PhysRevB.85.144403,PhysRevLett.102.177003,PhysRevLett.106.057004}.

We utilize a two-sublattice Holstein-Primakoff transformation
\begin{subequations}
\begin{eqnarray}
   S^z_i=S-a^\dag_i a_i,\\
   S^{+}_i=\sqrt{2S-a^{\dag_i} a_i}a_i \approx \sqrt{2S}a_i,\\
   S^{-}_i=a^\dag_i\sqrt{2S-a^\dag_i a_i}\approx\sqrt{2S}a^\dag_i,
\end{eqnarray}
\end{subequations}
with creation (annihilation) operators $a^\dag_i(a_i)$ to bosonize the Hamiltonian. The subscript $i=1, 2$ represents the sublattice indices as shown in figure~\ref{fig:1}(a). We then Fourier transform the bosonic operators
\begin{eqnarray}
 a_i=\frac{1}{\sqrt{N}}\sum_{\bf k} e^{i {\bf k}\cdot {\bf R_i }}a_{\bf k},  a^\dag_i=\frac{1}{\sqrt{N}}\sum_{\bf k} e^{-i {\bf k}\cdot {\bf R_i }}a^\dag_{\bf k},
\end{eqnarray}
to recast $\mathcal{H}$ in terms of the $a_{1,{\bf k}}$ and $a_{2,{\bf k}}$ bosons, where ${\bf k}$ is the wave vector in the BZ. In the Fourier transformed basis we define the operator $X^{\dag}_{{\bf k}}=(a^{\dag}_{1,{\bf k}},a^{\dag}_{2,{\bf k}},a_{1,-{\bf k}},a_{2,-{\bf k}})$ to represent the spin up and down sublattices. With this definition the LSWT Hamiltonian can be written as
\begin{eqnarray}
  \mathcal{H}&=&N e_c-\frac{1}{2}\sum_{\bf k} \textsf{Tr}\textsf{H}_{\bf k}+\frac{1}{2}\sum_{\bf k} X^{\dag}_{\bf k} \textsf{H}_{\bf k} X_{\bf k} \nonumber\\
  &=&Ne_0+\sum_k\sum^2_{n=1}\omega_{n,{\bf k}}b^{\dag}_{n,{\bf k}}b_{n,{\bf k}},
\end{eqnarray}
where $e_0=e_c-\frac{1}{2}\sum_{\bf k} \textsf{Tr}H_{\bf k}$ is the quantum zero-point energy correction, $e_c$ is  the classical ground-state energy, $\omega_{n,{\bf k}}$ is the LSWT dispersion, and
\begin{eqnarray}
\textsf{H}_{{\bf k}}=\left(\begin{array}{cccc}
          A_{{\bf k}} & B_{{\bf k}} &C_{{\bf k}} &D^{*}_{{\bf k}} \\
          B^{*}_{{\bf k}} & A_{{\bf k}} &D_{{\bf k}} &C_{{\bf k}} \\
          C_{{\bf k}} & D^{*}_{{\bf k}} &A_{{\bf k}} &B_{{\bf k}} \\
          D_{{\bf k}} & C_{{\bf k}} &B^{*}_{{\bf k}} &A_{{\bf k}}
          \end{array}\right),
\end{eqnarray}
with $A_{{\bf k}}=8KS^3+4J_{3}S+2J_{2}S\cos(k_x+k_y)$,$B_{\bf k}=(J_{1}S-2KS^3)(e^{ik_x}+e^{ik_y})$,
$C_{\bf k}=-2J_{2}S\cos(k_x+k_y)-2J_{3}S(\cos2k_x+\cos2k_y)$, $D_{\bf k}=-(J_{1}S+2KS^3)(e^{ik_x}+e^{ik_y})$.

General analytical expressions for the eigenvalues and eigenvectors can be obtained using standard paraunitary diagonalization schemes for a boson Hamiltonian \cite{PhysRevB.79.104421}. The eigenvalues of the two magnon bands are given by
\begin{subequations}
\begin{eqnarray}
\omega_{1,{\bf k}}&=(A^2+BB^{*}-C^2-DD^{*}\nonumber\\
            &-\sqrt{(4|AB-CD^{*}|^2-|B^{*}D^{*}-BD|^2)})^{\frac{1}{2}},\\
\omega_{2,{\bf k}}&=(A^2+BB^{*}-C^2-DD^{*}\nonumber\\
            &+\sqrt{ (4|AB-CD^{*}|^2-|B^{*}D^{*}-BD|^2)})^{\frac{1}{2}},
\end{eqnarray}
\end{subequations}
where $A\equiv A_{\bf k}$, $B\equiv B_{\bf k}$, and so on. The corresponding eigenvectors along with the normalization factor take the expressions $(U_1(\omega), V_1(\omega), \break U_2(\omega), V_2(\omega))$ and $N(\omega)$ respectively as
\begin{subequations}
\begin{eqnarray}
U_1(\omega)&=-(A+\omega)(A^2+BB^{*}-C^2-DD^{*}-{\omega}^2)\nonumber\\
         &+2ABB^{*}-C(B^{*}D^{*}+BD),\\
V_1(\omega)&=C(A^2+BB^{*}-C^2+DD^{*}-{\omega}^2)\nonumber\\
         &-A(B^{*}D^{*}+BD)-\omega(B^{*}D^{*}-BD),\\
U_2(\omega)&=B^{*}[(A+\omega)^2-BB^{*}+C^2]\nonumber\\
         &-2C(A+\omega)D+BD^2,\\
V_2(\omega)&=D(A^2+C^2-DD^{*}-{\omega}^2)\nonumber\\
         &+B^{*2}D^{*}-2AB^{*}C,\\
N(\omega)&=|U_1U_1^{*}-V_1V_1^{*}+U_2U_2^{*}-V_2V_2^{*}|.
\end{eqnarray}
\end{subequations}

Diagonalizing the system transforms it from the original $a_{1,{\bf k}}$ and $a_{2,{\bf k}}$ basis to the new Bogoliubov basis $b_{1,{\bf k}}$ and $b_{2,{\bf k}}$. The Bogoliubov transformation matrix $\textsf{S}$ which performs the diagonalization transformation matrix $\textsf{S}$ is given by
\begin{eqnarray}
\textsf{S}=\left(\begin{array}{cccc}
         \overline{U_1}(\omega_{1,k})&\overline{U_1}(\omega_{2,k})&\overline{V_1}(\omega_{1,k})&\overline{V_1}(\omega_{2,k})\\
         \overline{U_2}(\omega_{1,k})&\overline{U_2}(\omega_{2,k})&\overline{V_2}(\omega_{1,k})&\overline{V_2}(\omega_{2,k})\\
         \overline{V_1}(\omega_{1,k})&\overline{V_1}(\omega_{2,k})&\overline{U_1}(\omega_{1,k})&\overline{U_1}(\omega_{2,k})\\
         \overline{V_2}(\omega_{1,k})&\overline{V_2}(\omega_{2,k})&\overline{U_2}(\omega_{1,k})&\overline{U_2}(\omega_{2,k}),
         \end{array}\right)
\end{eqnarray}
where $1,k \equiv 1, {\bf k}$,  $2,k \equiv 2, {\bf k}$, $\textsf{S} \equiv \hat{S}_{\bf k}$, and we use the shorthand notation $\overline{U_1}(\omega)=U_1(\omega)/\sqrt{N(\omega)}$ to recast the eigenvectors in their normalized form. The new basis is defined as $X^{\prime} = \textsf{S}^{-1}X= (b_{1,{\bf k}},b_{2,{\bf k}},b^{\dag}_{1,-{\bf k}},b^{\dag}_{2,-{\bf k}})^{T}$.

\begin{figure}[t]
\centering
{\subfigure[Dispersion along the high symmetry points in the magnetic Brillouin Zone.]{
\includegraphics[scale=0.40]{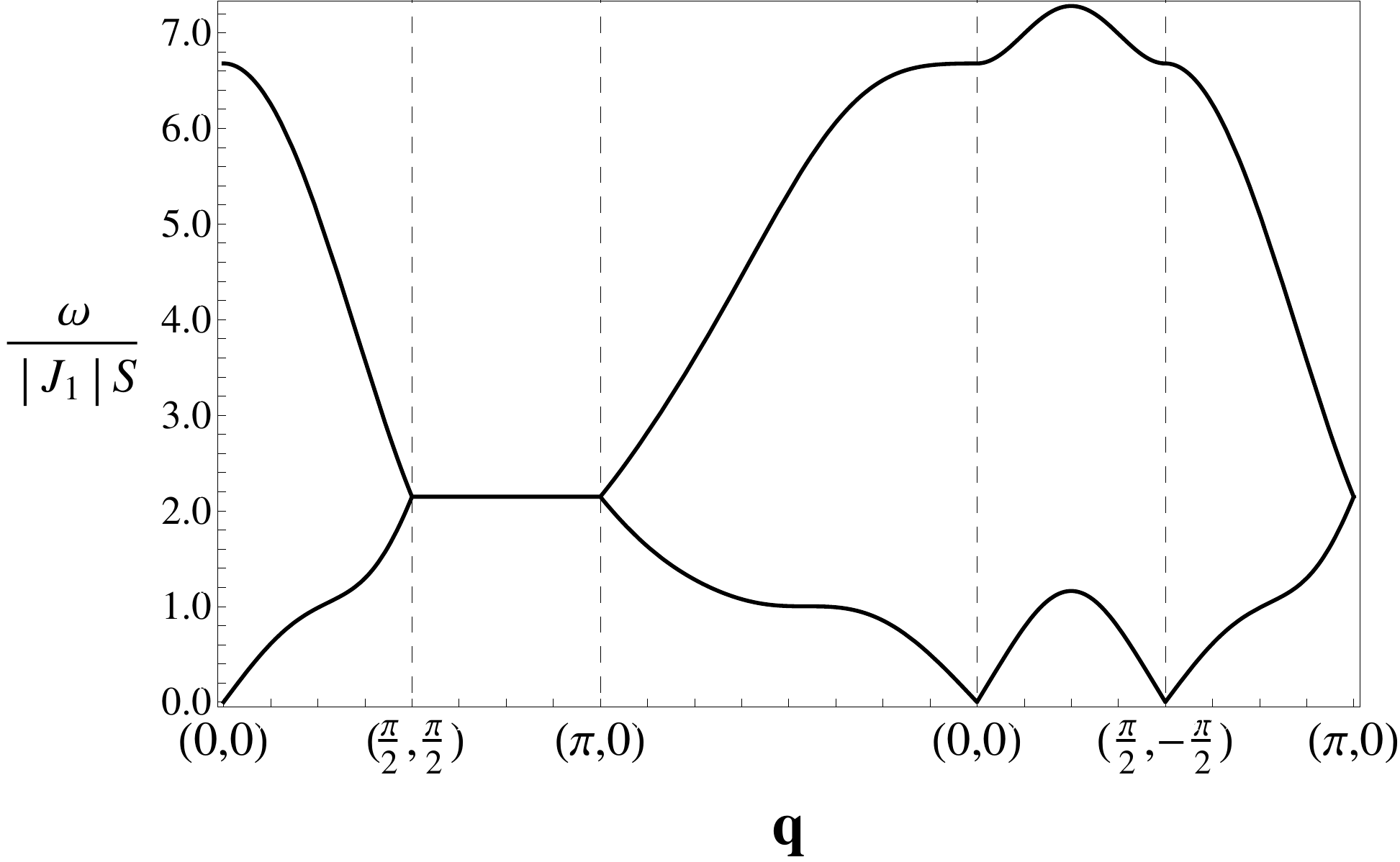}}}
{\subfigure[ 3D spin wave dispersion. ]{
\includegraphics[scale=0.40]{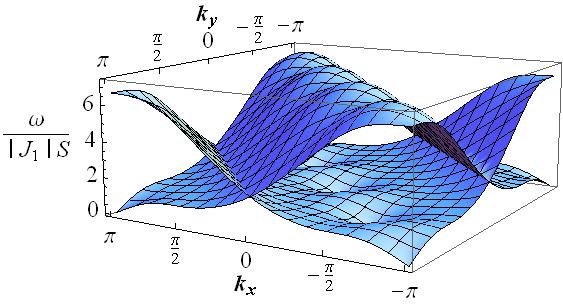}}}
\caption{Multi-band dispersion of the bicollinear antiferromagnetic (BCAF) model with exchange coupling
    parameters $J_1S=-1,J_2S=0.544,J_3S=0.28,KS^2=0.265~$\cite{PhysRevLett.106.057004}. S = 1 for all the plots.}
\label{Dispersion}
\end{figure}

The LSWT dispersion is displayed in figure~\ref{Dispersion}. We find that the dispersion of the BCAF has two magnon bands, a low energy acoustic branch and a high energy optical branch. The presence of the multiband feature makes the system more interesting for RIXS analysis since both intraband and interband transitions may be allowed. From the line plot in figure~\ref{Dispersion}(a) we observe that the acoustic and optical branches intersect at the $(\frac{\pi}{2},\frac{\pi}{2})$ point. Along the $(\frac{\pi}{2},\frac{\pi}{2}) \to (\pi,0)$ path the dispersions are degenerate and flat. But, tracking the dispersion along the
$(0,0)\to (\frac{\pi}{2},-\frac{\pi}{2})$ path we find that the optical and acoustic branches remain well separated without any degeneracy or intersection. The 3D spin wave dispersion in figure~\ref{Dispersion}(b) shows the separated acoustic and optical branch along the entire $(\pi,\pi) \to (\pi,-\pi)$.
\subsection{\label{dblcol}INDIRECT RIXS PROCESS}
\begin{figure}
\centering
\includegraphics[scale=0.45]{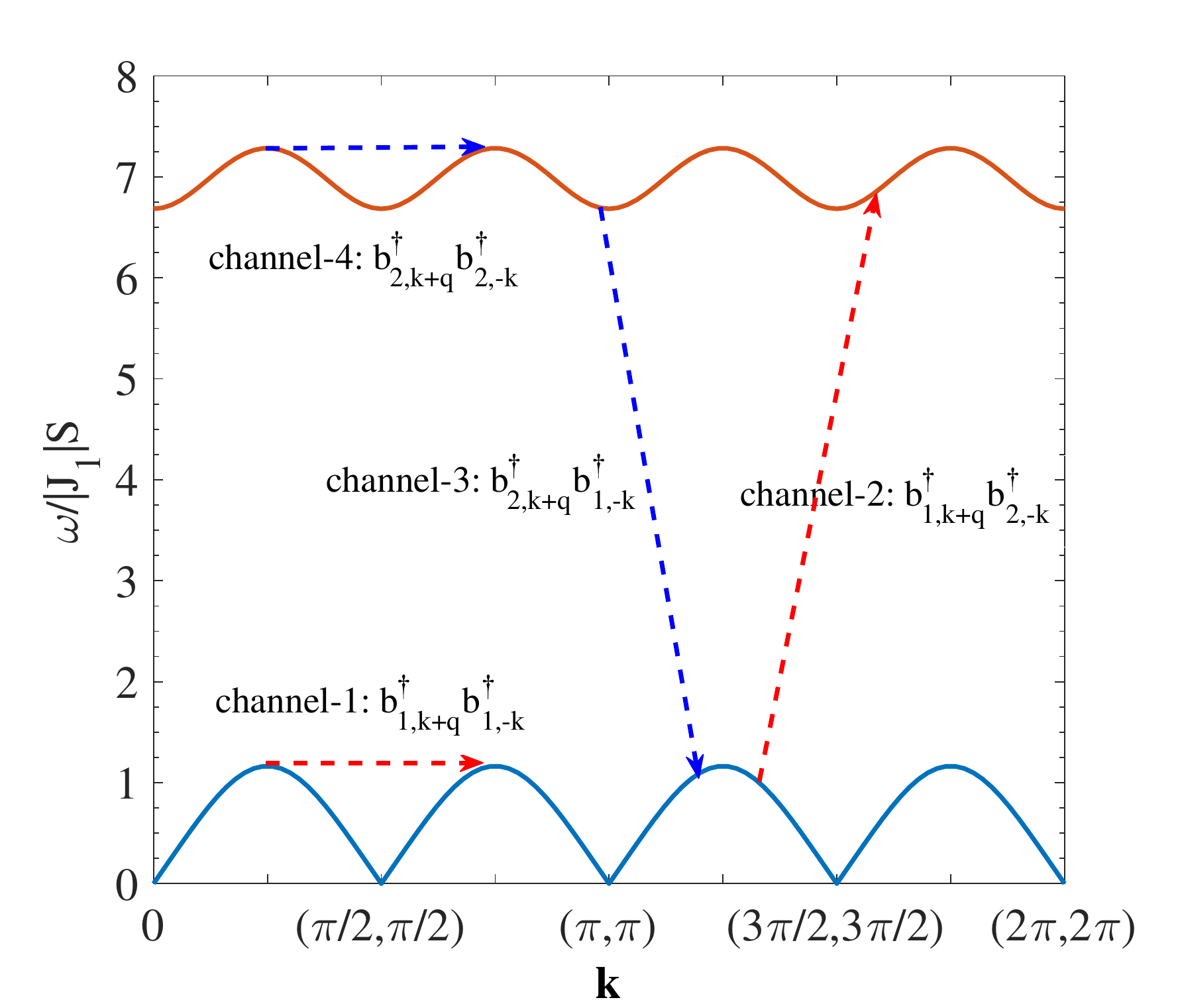}
\caption{Magnon dispersion bands and the associated bimagnon scattering channels. There are two intraband (channel-$1$ and channel-$4$) and two interband (channel-$2$ and channel-$3$) scattering processes . Channel-$1$ is the bimagnon excitation in the acoustic spin wave branch. Channel $4$ is the excitation created in the optical branch of the dispersion. The remaining channels $2$ and $3$ create bimagnon excitations which can transfer between the acoustic magnon and optical magnon branch.}
\label{fig:channels}
\end{figure}

In the {\it indirect} K-edge RIXS process, an inner shell 1$s$ electron is promoted to an empty 4$p$ band by absorbing a photon. A core hole shakes up of the system in the intermediate state can generate bimagnon excitations. Assuming that the intermediate state is short lived (typically of the order of femtoseconds), we can employ the well established ultrashort core-hole lifetime (UCL) expansion approximation \cite{EuroPhysLett.73.121,PhysRevB.75.115118}. Within this approximation to the lowest order, the bimagnon RIXS scattering operator is defined as\cite{PhysRevB.83.245133}
\begin{eqnarray}
\label{eq:rixsop}
   \hat{O}({\bf q})=\sum_{i,j}J_{i,j}e^{i{\bf q}\cdot {\bf R_{i}}}S_{i}\cdot S_{j},
\end{eqnarray}
where the $i,j$ runs over the NN, NNN, and NNNN lattice points. To compute the bimagnon RIXS spectrum we first transform the RIXS operator to its bosonized form using the two-sublattice Holstein-Primakoff transformation. Second, we perform a Bogoliubov transformation. In the transformed basis the RIXS scattering operator can be expressed as
\begin{eqnarray}
 \hat{O}({\bf q})=\sum_{\bf k} X^{\prime\dag}_{\bf k+q}\hat{O}^\prime_{\bf k}({\bf q}) X^\prime_{\bf k},
\end{eqnarray}
where $X^{\prime\dag}_{\bf k+q}=(b^\dag_{1,{\bf {k+q}}},b^\dag_{2,{\bf k+q}},b_{1,{\bf-k-q}},b_{2,{\bf -k-q}})$ and
$X^{\prime}_{\bf k}=(b_{1,{\bf k}},b_{2,{\bf k}},b^{\dag}_{1,-{\bf k}},b^{\dag}_{2,-{\bf k}})^{T}$. The Bogoliubov transformed bimagnon RIXS operator is thus defined as
\begin{eqnarray}
  \hat{O}^\prime_{\bf k}({\bf q})&=&\left(\begin{array}{cc}
                                \hat{O}^\prime_{11} & \hat{O}^\prime_{12} \\
                                \hat{O}^\prime_{21} & \hat{O}^\prime_{22}\\
                              \end{array}\right),\nonumber\\
                      \hat{O}^\prime_{\bf k}({\bf q})        & =&\hat{S}^\dag_{\bf k+q}\left(\begin{array}{cc}
                                \hat{O}_{11} & \hat{O}_{12} \\
                                \hat{O}_{21} & \hat{O}_{22}\\
                              \end{array}\right)\hat{S}_{\bf k}.
\end{eqnarray}
The expressions for the individual diagonal and off-diagonal components are given by
\begin{eqnarray}
\hat{O}_{11}&=&\left(\begin{array}{cc}
                          A_{\bf 1k}               &B_{\bf k+q}+B_{\bf k}        \\
                          B^{*}_{\bf k+q}+B^{*}_{\bf k}     &A_{\bf 2k}\\
                          \end{array}\right)=\hat{O}_{22},\\
\hat{O}_{12}&=&\left(\begin{array}{cc}
                          C_{\bf k+q}+C_{\bf k}               &D^{*}_{\bf k+q}+D^{*}_{\bf k}        \\
                          D_{\bf k+q}+D_{\bf k}     &C_{\bf k+q}+C_{\bf k}\\
                          \end{array}\right)=\hat{O}_{21},\\
\end{eqnarray}
with $A_{\bf 1k}=8KS^3+2J_{2}S\cos(k_x+k_y)+2J_{2}S\cos(k_x+q_x+k_y+q_y)-2J_{2}S\cos(q_x+q_y)+2J_{2}S\cos(q_x-q_y)
+4J_{3}S+2J_{3}S(\cos2q_x+\cos2q_y)+(J_{1}S+2KS^3)(e^{-iq_x}+e^{iq_y})-(J_{1}S-2KS^3)(e^{iq_x}+e^{-iq_y}),
A_{1{\bf k}}=A^{*}_{2{\bf k}}$. After the Bogoliubov transformation, the off-diagonal RIXS operator components have the form
\begin{eqnarray}
  \hat{O}^\prime_{12}=\left(\begin{array}{cc}
                                M_{11} & M_{12} \\
                                M_{21} & M_{22}
                              \end{array}\right),
\end{eqnarray}
where $M_{ij}$ ($i,j = 1, 2$) are the RIXS intraband and interband transition channels in the transformed basis. In figure~\ref{fig:channels} we show the scattering process corresponding to these four channels. Thus, the entire  {\it indirect} K-edge bimagnon RIXS operator $\hat{O}_2({\bf q})$ can be constructed out of the $2\times2$ block matrices $\hat{O}^\prime_{12}$ as
\begin{eqnarray}
\label{eq:totrixs}
  \hat{O}_2({\bf q})&=\sum_{\bf k}( M_{11}b^\dag_{1,{\bf k+q}}b^\dag_{1,{\bf -k}}+M_{12}b^\dag_{1,{\bf k+q}}b^\dag_{2,{\bf -k}}\nonumber\\
  &+M_{21}b^\dag_{2,{\bf k+q}}b^\dag_{1,{\bf -k}}+M_{22}b^\dag_{2,{\bf k+q}}b^\dag_{2,{\bf -k}}+h.c).
\end{eqnarray}
The excitations corresponding to the $M_{11}$ and $M_{22}$ channels are classified as intraband. The other two, $M_{12}$ and $M_{21},  $are the interband channels. Typically, the RIXS operator for one magnon band systems such as the J$_{1}$-J$_{2}$ Heisenberg model has a 2 $\times$ 2 matrix form~\cite{PhysRevB.89.165103}. However, for this multiband problem it is a 4 $\times$ 4 matrix. The frequency and momentum-dependent bimagnon scattering intensity is then given by
\begin{eqnarray}
\mathcal{I}({\bf q},\omega)&\propto\sum_f|\bra{i}\hat{O}_2({\bf q})\ket{f}|^2\delta(\omega-\omega_{fi})\nonumber\\
&=-\frac{1}{\pi}\mathrm{Im}\mathrm{G}({\bf q},\omega),
\end{eqnarray}
where $\ket{i}$ and $\ket{f}$ are the initial and final states with corresponding transfered energy $\omega_{fi}$ and momentum ${\bf q}$, respectively.
The time-ordered correlation function is given by
\begin{equation}
  \mathrm{G}({\bf q},\omega)=-i\int_0^\infty\mathrm{d}t\ e^{i\omega t}\bra{i}\mathcal{T}\hat{O}^\dag_2({\bf q})(t)\hat{O}_2({\bf q})(0)\ket{i}.
\end{equation}
The momentum-dependent bimagnon Green's function defined as
\begin{eqnarray}
&\Pi_{ij}({\bf q},t;{\bf k,k'})\nonumber\\
&=-i\bra{0}\mathcal{T}b_{i,{\bf k+q}}(t)b_{j,{\bf k}}(t)b_{i,{\bf k^\prime+q}}^\dag(0)b_{j,{\bf k^\prime}}^\dag(0)\ket{0},
\end{eqnarray}
can be expanded in terms of the one magnon propagators
\begin{subequations}
\begin{eqnarray}
\mathrm{G}_{b_1,b_1}({\bf k},t)=-i\bra{0}\mathcal{T}b_{1,{\bf k}}(t)b_{1,{\bf k}}^\dag(0)\ket{0},\\
\mathrm{G}_{b_2,b_2}({\bf k},t)=-i\bra{0}\mathcal{T}b_{2,{\bf k}}(t)b^\dag_{2,{\bf k}}(0)\ket{0},
\end{eqnarray}
\end{subequations}
for the $b_1$ and $b_2$ magnons,
where $\mathcal{T}$ is the time ordering operator, and $\ket{0}$ is the ground state.
The LSWT bimagnon propagator is denoted by $\Pi_0({\bf q},\omega;{\bf k})$. The RIXS intensity spectrum, expressed in terms of the intra- and inter- band channel transition strengths, is given by
\begin{eqnarray}
\label{eq:rixsinten}
 & \mathcal{I}({\bf q},\omega)\propto-\frac{1}{\pi}\mathrm{Im}\left[\sum^{1,2}_{i,j}|M_{ij}|^2\Pi_{i,j}({\bf q},t;{\bf k,k'})\right]\\
                       &=-\frac{1}{\pi}\mathrm{Im}\left[\sum^{1,2}_{i,j}|M_{ij}|^2\frac{1}{\omega-\omega_{i,{\bf k+q}}-\omega_{j,{\bf -k}}+i0^{+}} \right].
\end{eqnarray}

\section{Results and discussion}
\subsection{RIXS intensity spectrum}\label{Sec:intensity}
\begin{figure}[t]
\centering
\includegraphics[scale=0.5]{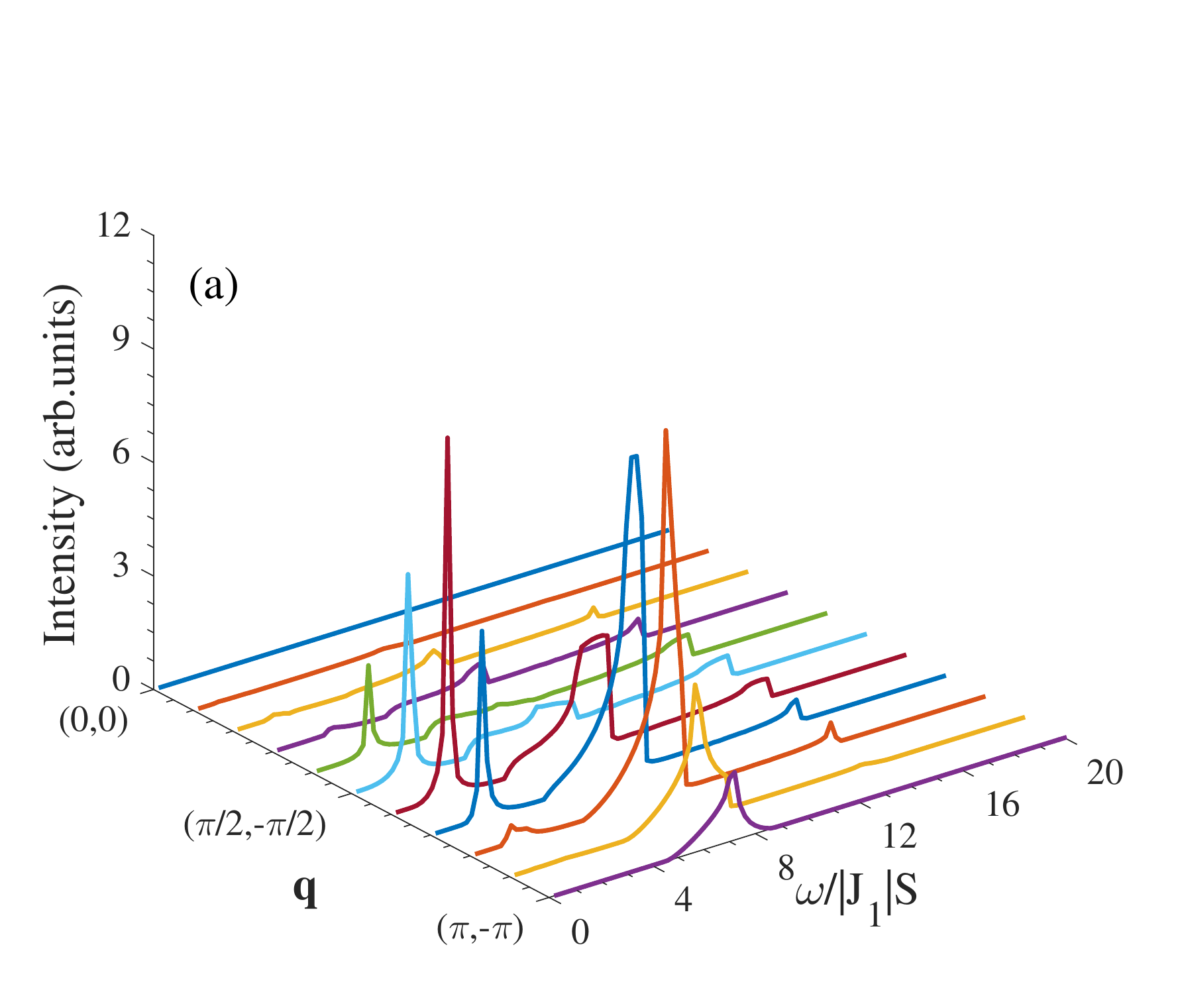}
\includegraphics[scale=0.5]{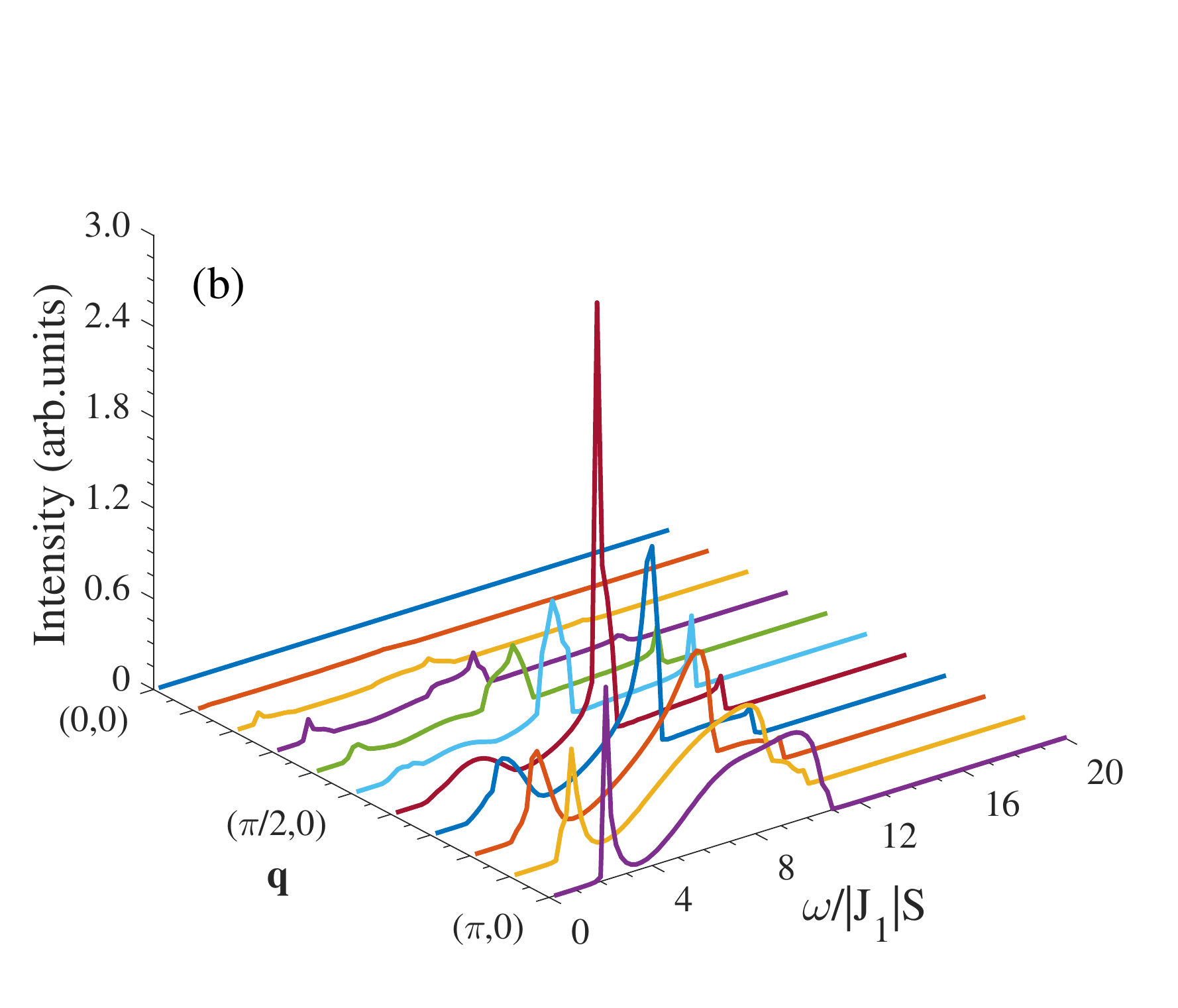}
\caption{Energy - momentum dependence of the {\it indirect} K$-$edge RIXS spectrum of the multi spin wave band $\alpha$-FeTe chalcogenide system.  (a) Momentum ${\bf q}$ path along $(0,0)\to (\pi,-\pi)$. (b) Momentum $q$ path along $(0,0)\to (\pi,0)$. Parameter choice for $\alpha$-FeTe  $J_1S=-1,J_2S=0.544,J_3S=0.28,KS^2=0.265$ . A ferromagnetic (FM) $J_1$ and an AF $J_3$ must be included while the $J_2$ does not differ significantly from ferrochalcogenides in $\alpha$-FeTe system. The BCAF ordering is stable when $2J_3>J_2$ and $2J_2>|J_1|$ \cite{ PhysRevLett.102.177003, PhysRevLett.106.057004, EuroPhysLett.86.67005,PhysRevB.85.144403}.}
\label{fig:4}
\end{figure}

We use equation~(\ref{eq:rixsinten}) to compute the frequency and momentum dependent {\it indirect} K-edge bimagnon RIXS intensity spectra $\mathcal{I}({\bf q},\omega)$ of the BCAF model. In figure~\ref{fig:4} we show the evolution of the spectra along two momentum paths. For the first path, see figure~\ref{fig:4}(a), we display the spectrum along the $(0,0)\to (\pi,-\pi)$ path which includes the BCAF magnetic ordering wave vector $(\pi/2,-\pi/2)$. In contrast to the square lattice models where the spectrum is known to vanish at the magnetic ordering wave vector, in the BCAF system the RIXS intensity is \emph{non-zero} at $(\pi/2,-\pi/2)$. But, similar to the Heisenberg square lattice {\it indirect} K-edge RIXS spectrum we find that the intensity is zero at ${\bf q}=(0,0)$~\cite{EuroPhysLett.80.47003,PhysRevB.75.214414,PhysRevB.89.165103}. Away from the Brillouin zone edge the X-ray spectrum starts to develop non-zero contributions from the four different momentum channels. The channel 1 intraband contribution grows in strength reaching its peak value at ${\bf q} \approx (3\pi/5,0)$ before decaying to zero at ${\bf q} = (\pi,-\pi)$. The channel 4 intraband signal is relatively weak along this path. Between $(\pi/2,-\pi/2)$ and $(\pi,-\pi)$ the major contribution to the RIXS signal is from the interband scattering processes with the maximum intensity at ${\bf q} \approx (4\pi/5,0)$. In most cases we observe a two-peak or a single peak structure in the spectrum.

In figure~\ref{fig:4}(b) we show our results for the second path along $(0,0)\to (\pi,0)$. Similar to the previous path,the $(0,0)$ point has a vanishing signal in where all the intraband and interband RIXS excitations are suppressed. For transfer momentum ${\bf q}$ approximately beyond $(2\pi/5,0)$ ,the RIXS intensity peaks are prominent and can be easily distinguished. The maximum intensity occurs for the interband channel at ${\bf q} \approx (3\pi/5,0)$. In most cases we observe three rather than four peaks because the interband channels are degenerate.  It is also possible to have a two-peak or a single peak structure under suitable conditions. With increasing momentum ${\bf q}$ ,the optical branch response undergoes a spectral downshift. This causes the RIXS response from the interband and intraband fourth channel to overlap, thereby creating a broad shoulder.

\begin{figure}
\centering
\includegraphics[scale=0.5]{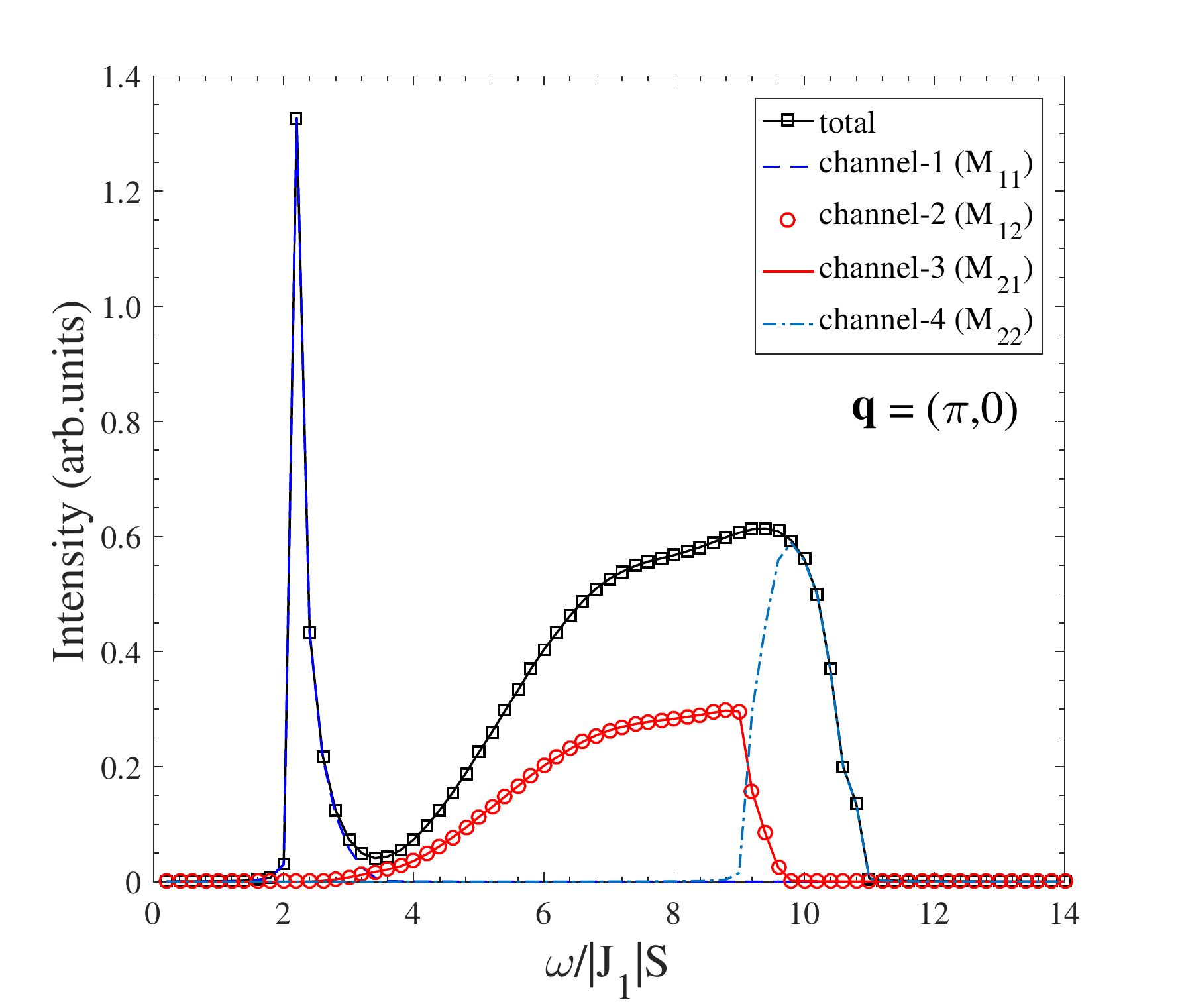}
\caption{Individual and total K-edge bimagnon RIXS intensity of the BCAF model. Parameters are same as in Figure~\ref{fig:4}. The intraband RIXS channels $(M_{11}, M_{22})$ correspond to distinct RIXS peaks. However, the interband channels $(M_{21}, M_{12})$ are degenerate.}
\label{fig:5}
\end{figure}

To understand the features of the RIXS spectra discussed above, we should analyze the RIXS spectra from the individual intraband and interband  channels. In figure~\ref{fig:5} we plot the individual intensity from the four LSWT RIXS channels $M_{ij}$ at ${\bf q}=(\pi,0)$. Channel 1 excitations are exclusively in the acoustic branch. Channels $2$ and $3$  are composed of transitions between an acoustic magnon and an optical one. Channel $4$ corresponds to the optical magnons. The acoustic bimagnon excitation spectrum is the lowest energy response of the spectrum and the optical one is the highest. As shown in figure~\ref{fig:5} the spectra from channels $2$ and $3$ are degenerate. The degeneracy of the middle two spectra explain the occurrence of a three peak and not a four peak structure in the RIXS spectrum of figure~\ref{fig:4}(b). The sum  of all the individual contributions give rise to a spectrum that has one sharp peak and one broadened peak (black solid line with squares).

\begin{figure}
\centering
\includegraphics[scale=0.5]{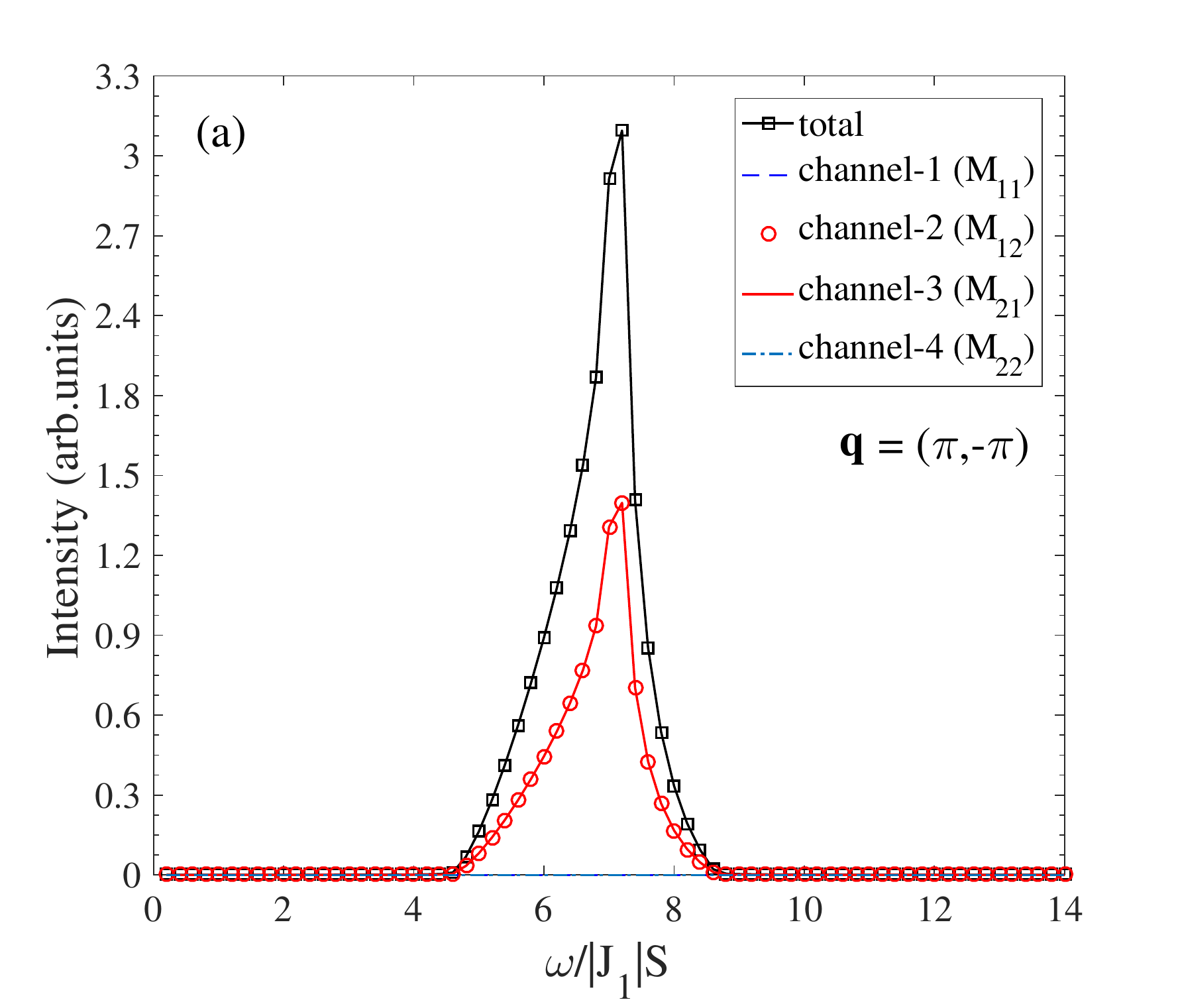}
\includegraphics[scale=0.5]{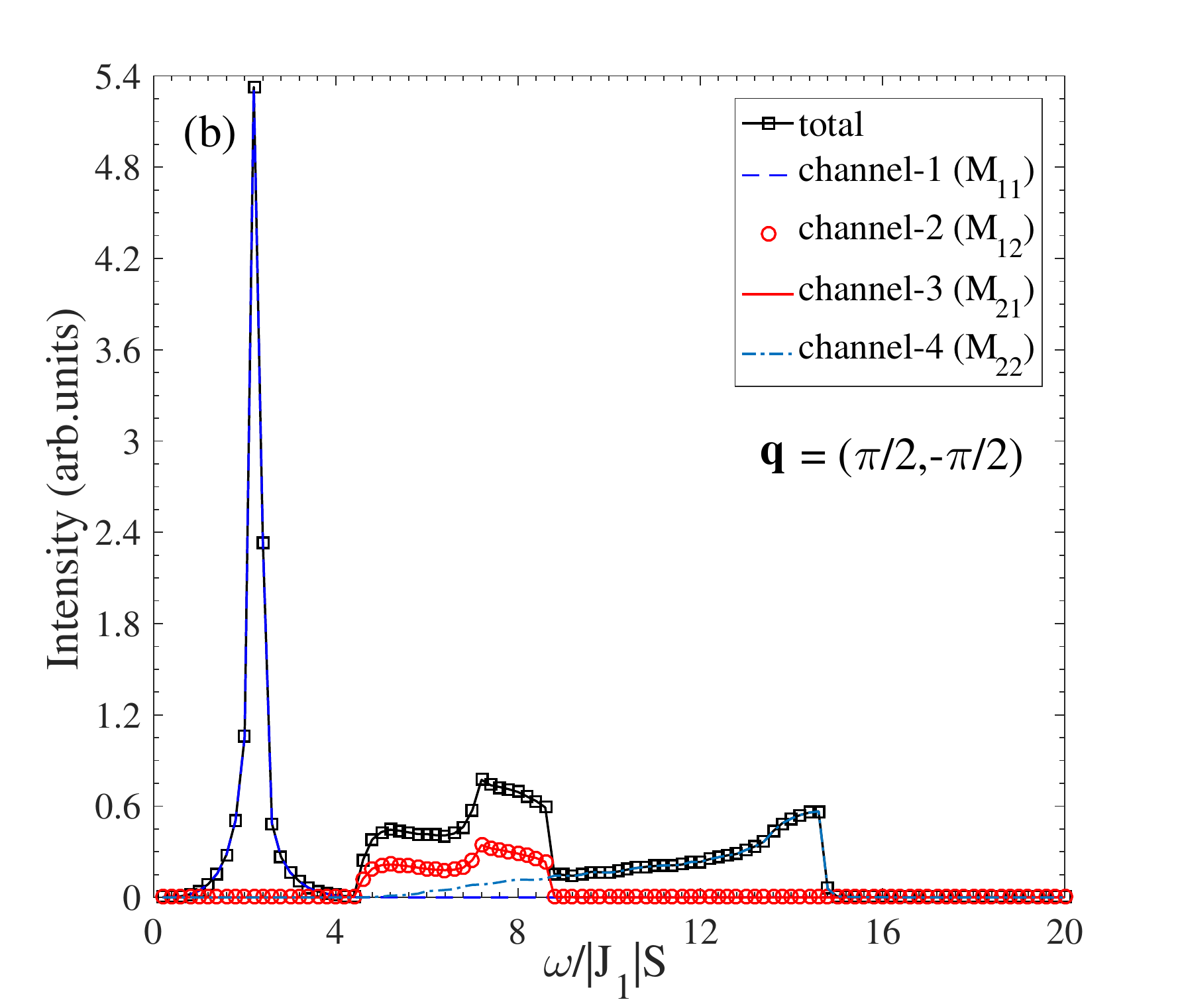}
\caption{RIXS spectra comparison at two different scattering momentum in the magnetic Brillouin zone.}
\label{fig:6}
\end{figure}

Our previous discussion of the RIXS spectra along the $(0,0)\to (\pi,-\pi)$ path pointed to the existence of a single peak structure. It is useful to study figure~\ref{fig:6}(a) to understand the origin of the single peak. We observe in figure~\ref{fig:6} that the intraband contributions are identically zero at this transfer momentum ${\bf q}=(\pi,-\pi)$ . However, the degenerate interband signals are not. Thus we have a finite contribution giving rise to the single peak spectrum. A physical explanation of the origins of this peak can be understood by inspecting the arrangement of the spins. Along the (1, -1) direction in the real space lattice, highlighted by the dashed blue box in figure~\ref{fig:1}(a), the spins are coupled antiferromagnetically and belong to the same sublattice (1 or 2). We know from previous studies on the square antiferromagnetic lattice that the RIXS response vanishes at the $(\pi,-\pi)$ point~\cite{EuroPhysLett.80.47003,PhysRevB.75.214414,PhysRevB.89.165103}. Thus, the contributions from the two intraband channels are identically zero. However, the BCAF lattice is composed of staggered AF chains coupled along the diagonal by interactions. These interactions created interband magnetic excitations giving rise to channels 2 and 3 in the RIXS spectrum. Thus, the interband RIXS intensity is not zero. The RIXS spectra should also be non-zero for the ferromagnetically coupled spins along the other diagonal direction in real space. Along this path the system is a set of staggered ferromagnetically coupled chains. For a pure ferromagnetic system the RIXS response is known to be zero. However, here it is not (figure not shown).

Figure~\ref{fig:6}(b) shows the detailed channel contributions of the non-vanishing RIXS spectrum at the ordering wave vector. First, using equation (\ref{eq:rixsop}) at the magnetic ordering wave vector ${\bf q}=(\pi/2,-\pi/2)$ we have $e^{i{\bf q}\cdot {\bf R}_i}=-i$ if ${\bf R}_{i}$ is in sublattice 1 and $e^{i{\bf q}\cdot {\bf R}_i}=1$ if ${\bf R}_{i}$ is in sublattice 2 (assuming that at ${\bf R}_{i}=(0,0)$ , we are in sublattice 1). Thus, we find the RIXS operator
\begin{eqnarray}
\hat{O}_{{\bf q}=(\frac{\pi}{2},-\frac{\pi}{2})}=-i\sum_{i\in 1 ,j}J_{i,j}{\bf S}_i\cdot{\bf S}_j+\sum_{i\in 2 ,j}J_{i,j}{\bf S}_i\cdot {\bf S}_j,
\end{eqnarray}
does cannot cancel when applied to an initial state which is non-symmetric under the interchange of the sublattices. Hence, the RIXS intensity is \emph{non-zero} at the ordering wave vector.

\begin{figure}
\centering
\includegraphics[scale=0.5]{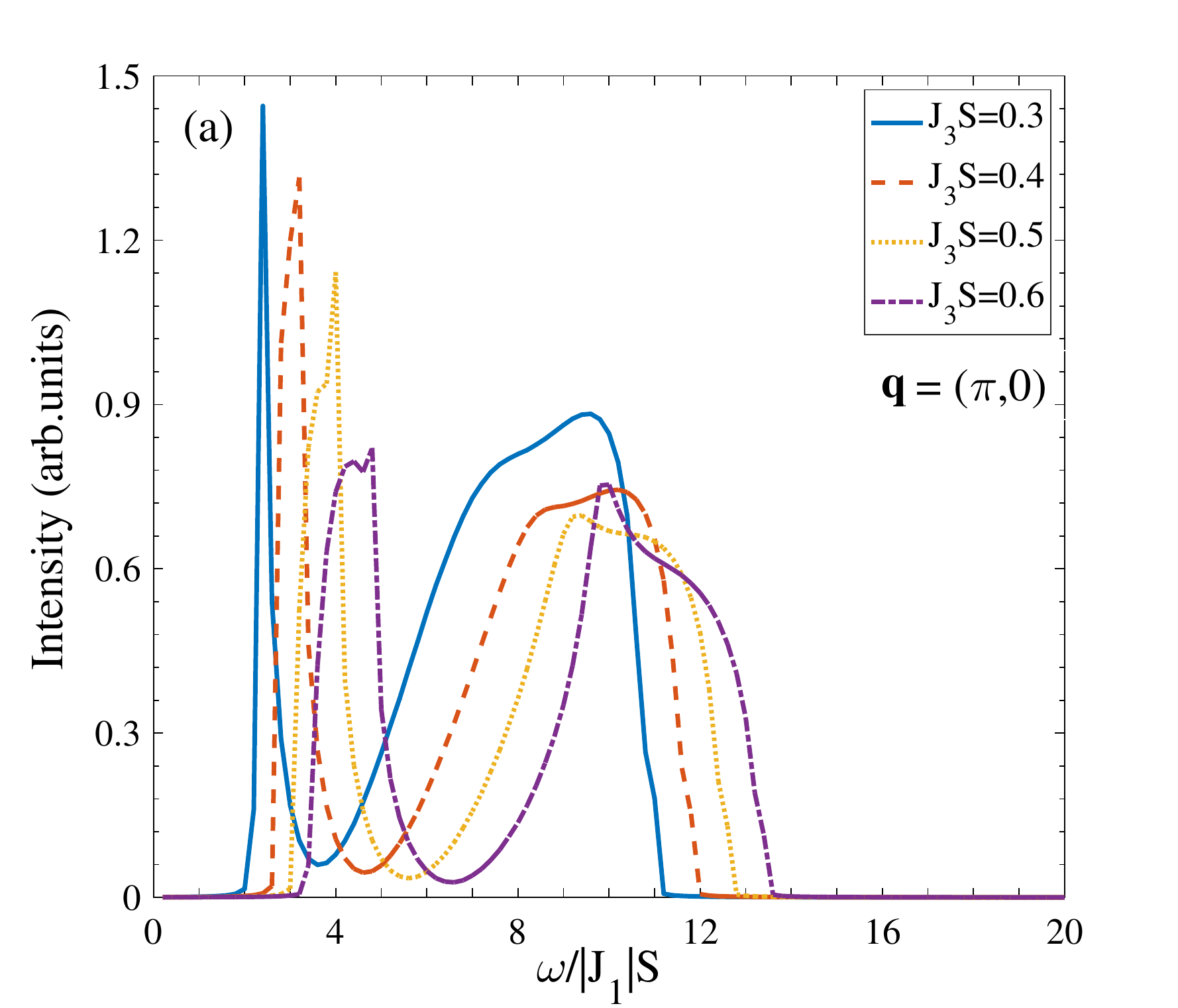}
\includegraphics[scale=0.5]{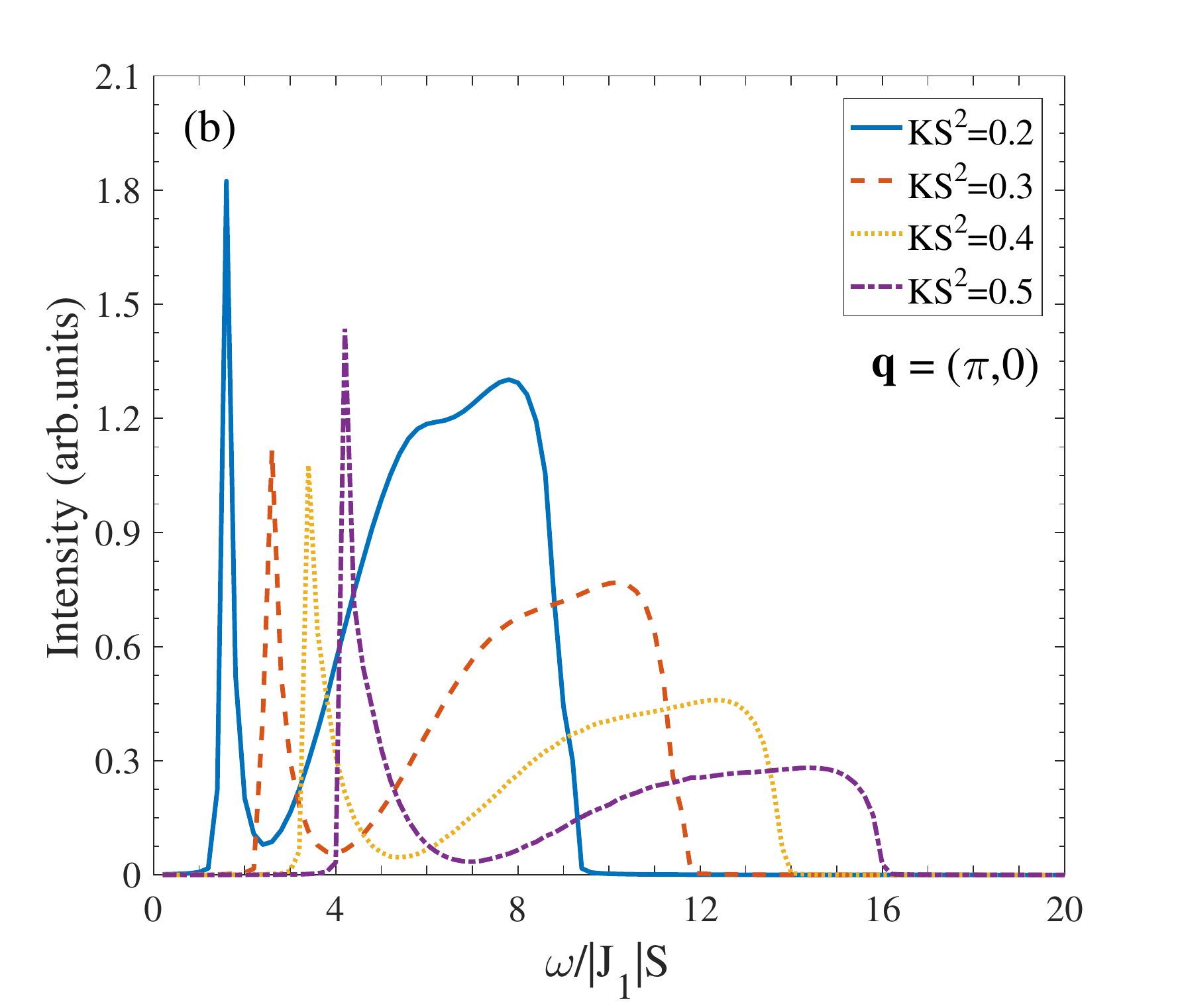}
\caption{(a)spectras of different $J_3$.($J_1S=-1$,$J_2S=0.544$,$KS^2=0.265$).(b)Position of peak with different $KS^2$.($J_1S=-1$,$J_2S=0.544$,$J_3S=0.28$)}
\label{fig:7}
\end{figure}

We compute the RIXS spectra for different $J_{3}$ and $K$ to understand the consequences of longer-range and more complex interactions. From figure~\ref{fig:7}
 we observe that the RIXS spectrum is sensitive to variations in both the $J_{3}$ and $K$ interactions. Overall the spectrum is pushed towards higher energies. There is also a spectral intensity reshuffling from the low energy acoustic bimagnon excitation branch to the high energy optical branch. This is evident from the bulge that is generated as $J_{3}$ is increased in strength. With increasing $J_{3}$ coupling the bandwidth of the acoustic and optical branch is enhanced. This increase is evident from the shift towards higher for both the intraband and interband RIXS response (within LSWT). However, increasing the biquadratic coupling affects the optical branch more than the acoustic. For large K values, the optical branch becomes almost dispersionless (nearly a flat band). The acoustic branch is also shifted towards higher energy values. The increased energy causes the entire RIXS spectrum to  be shifted towards higher energy (within LSWT). However, the intensity of the fourth interaction channel is progressively suppressed because the optical branch becomes narrower. The interband interactions are also affected. For the $\alpha$-FeTe materials the biquadratic coupling $K$ is typically considered to be nonzero~\cite{PhysRevB.85.134406}. Thus the effect of the biquadratic coupling on the dispersion is more
drastic than the J$_{3}$ coupling. Based on our calculations, we find that the RIXS spectrum is more sensitive to changes in K compared to J$_{3}$.
\subsection{Conclusion}
Using LSWT, we calculated the K-edge {\it indirect} bimagnon RIXS spectrum for the $\alpha$-FeTe system. The underlying BCAF magnetic ordering, modeled by a
$J_1-J_2-J_3-K$ Heisenberg Hamiltonian, introduces two magnon bands. The presence of low energy acoustic and high energy optical branches create additional RIXS channels (within LSWT) which are typically absent in the square lattice Heisenberg model case. Thus, we explore the RIXS response from a multi-channel perspective. Specifically, we find four channels of scattering, two of which are classified as intraband and the other two as interband. Calculating the RIXS intensity for various paths along the magnetic Brillouin zone yields some interesting conclusions. We find that, in contrast to the standard $J_{x}-J_{y}-J_{2}$ RIXS spectrum, the intensity for the BCAF model does not vanish at the BCAF magnetic ordering wave vector ${\bf q } = (\pi/2,-\pi/2)$. The spectra display either a single, double, or three peak structure at appropriate transfer momentum ${\bf q}$. The occurrence of the single peak at the $(\pi,-\pi)$ point is physically interesting. The BCAF magnetic ordering can be viewed as a collection of antiferromagnetically or ferromagnetically coupled strips along the  $(1,-1)$ or $(1,1)$ direction respectively. While ordinarily the response of an AF or ferromagnetic ordering would give zero RIXS intensity, in the BCAF model the presence of two sublattices makes the spectrum non-zero at the magnetic ordering wave vector. While experimentally neutron scattering can be used to clearly identify the BCAF ordering, in situations where there is possible ambiguity we propose that the characteristic features of the K-edge bimagnon RIXS intensity can help to resolve this issue. We also study the effects of $J_{3}$  and the $K$ coupling for the model. These interactions do affect the RIXS spectrum even at the LSWT level. Note, the large S value of the $\alpha$-FeTe system allows us to perform the calculation simply within the LSWT approximation without having to incorporate the effects of quantum fluctuations. Even if interactions were included beyond the LSWT level, the overall results would be mildly modified, since for large-S values quantum fluctuations have minimal effects~\cite{PhysRevB.89.165103}. As beamlines of enhanced resolution are deployed globally, we hope our investigation in this paper will inspire experimentalists to study the RIXS spectrum of this  important chalcogenide system.

\begin{acknowledgments}
T.D. acknowledges invitation, hospitality, and kind support from Sun Yat-Sen University. T. D. also acknowledges funding support from Augusta University Scholarly Activity Award. Z. H. and D. X. Y. are support by NSFC-11574404, NSFC-11275279, NSFG-2015A030313176, National Key Research and Development Program (Grant No. 2017YFA0206203), Special Program for Applied Research on Super Computation of the NSFC-Guangdong Joint Fund, Leading Talent Program of Guangdong Special Projects. S. M. acknowledges funding support from the Augusta University CURS Summer Scholars program. 
\end{acknowledgments}

\bibliographystyle{apsrev4-1}
\newpage
\bibliography{bibfile}
\end{document}